\documentclass[aps,twocolumn,notitlepage,prb,10pt,showpacs, noeprint]{revtex4-1}
\usepackage{bm}
\usepackage{textcomp}
\usepackage{hyperref}
\usepackage{url}
\usepackage{amsmath}
\usepackage{amssymb}
\usepackage{amsxtra}
\usepackage{amscd}
\usepackage{amsthm}
\usepackage{amsfonts}
\usepackage{eucal}
\usepackage{latexsym,amsthm}
\usepackage{hhline}
\usepackage{multirow}
\usepackage{pstricks}
\usepackage{color}
\usepackage{graphicx}
\usepackage{enumitem}

\newcommand{\nc}{\newcommand}
\nc{\on}{\operatorname}
\nc{\wt}{\widetilde}
\nc{\Wick}{{\mathbb :}}
\nc{\R}{{\mathbb R}}

\newcommand{\beq}{\begin{equation}}
\newcommand{\eeq}{\end{equation}}
\newcommand{\bmul}{\begin{multline}}
\newcommand{\emul}{{\end{multline}}}
\newcommand\beqa{\begin{eqnarray}}
\newcommand\eeqa{\end{eqnarray}}
\newcommand\bea{\begin{array}}
\newcommand\eea{\end{array}}
\newcommand\ba{\begin{array}}
\newcommand\ea{\end{array}}
\newcommand{\nn}{\nonumber}

\newcommand{\neqa}{\nonumber\end{eqnarray}}

\newcommand{\eq}[1]{Eq. (\ref{#1})}

\newcommand{\Eq}[1]{Eq. (\ref{#1})}
\newcommand{\Eqs}[2]{Eqs. (\ref{#1}) and (\ref{#2})}

\renewcommand{\d}{\partial}

\renewcommand{\L}{{\cal L}}

\nc{\CH}{{\mathcal H}}
\nc{\Db}{{\bar D}}
\nc\comment[1]{}

\nc{\CM}{{\mathcal M}}
\nc{\CN}{{\mathcal N}}

\newcommand{\re}{\relax{\rm I\kern-.18em R}}

\nc{\meV}{{\mathrm{\,meV}}}
\nc{\cG}{{\mathcal G}}

\renewcommand{\bar}{\overline} 

\def\eV{{\mathrm{eV}}}
\nc{\al}{{\alpha}}

\def\nm{{\,{\rm nm}}}

\def\eps{{\epsilon}}
\def\sign{{\rm \, sign }}

\renewcommand{\)}{\right)}

\begin{document}
\title{Suppressed compressibility of quantum Hall effect edge states in epitaxial graphene on SiC}
\author{Sergey Slizovskiy}
\email{on leave of absence from NRC ``Kurchatov Institute'' PNPI, Russia.}
\author{Vladimir I. Fal'ko}
\affiliation{National Graphene Institute, The University of Manchester, M13 9PL, Booth st. E.,  Manchester, UK }
\keywords{graphene, Quantum Hall Effect, edge states, Quantum Resistance Standard, metrology }
\pacs{}
\begin{abstract}
We determine conditions for the formation of compressible stripes near the quantum Hall effect (QHE) edges of top-gated epitaxial graphene on Si-terminated SiC (G/SiC)
and compare those to  graphene exfoliated onto insulating substrate in the field-effect-transistor (GraFET) geometry.
 For G/SiC, a  large density of localised surface states on SiC just underneath graphene layer and  charge transfer between them lead both to doping of graphene  and to screening of potential profile near its edge.
This suppresses formation of compressible stripes near QHE edges in graphene, making them much narrower than the corresponding compressible stripes in GraFETs. 
\end{abstract}

\maketitle

\section{Introduction}
Edge states in the quantum Hall effect (QHE) systems \cite{vonKlitzing80,Laughlin81,EdgeThouless2,EdgeThouless1, ButtikerChiral,vonKlitzing86}
are chiral, providing transport channels that carry electrons along the edge in the direction set by magnetic field polarity. 
Over the years,   investigations of  
the structure of edge states in semiconductor heterostructures were carried out for the understanding  of  current noise\cite{EdgeEquilibration, MartinFeng90, GerhardsCurrents, PartialEquilibration09, Kovrizhin}
and cooling rates\cite{2DEGHeat,GrapheneExperiment, SSFalko} in the QHE regime. 
For electrons in GaAs/AlGaAs devices, it has been shown\cite{Gerhards88,Shklovskii92, Chalker93,Fogler94,GerhardsCurrents} that, 
in a strong magnetic field, an electrostatically soft edge 
of a 2D electron gas
reconstructs   into a 
sequence of compressible and incompressible stripes.
A similar possibility was  recently suggested  for graphene \cite{Efetov, EdgeSiO2, IvanEdgePotential, StripeHubbard,CompressibleInAStripe}, where an essential difference 
arises from 
a $|y|^{-1/2}$ singularity in the charge density near the edge\footnote{
In contrast to 2DEG created by smooth gate potentials, graphene has a finite density of states up to the very edge of the flake. This leads to 
  a sigularity in the charge density near the edge, known to appear in thin charged metal plates \cite{TextBook}}
graphene-specific 
edge states\cite{EdgeStates1, EdgeStates2, Abanin2007, EdgeStates3, AbaninEdgeWithScreening}. 

When epitaxial graphene is grown on Si-terminated face of SiC (G/SiC), a ``dead layer'' of carbons forms on the SiC surface, right underneath graphene\cite{EpitaxialGrowth, ReviewBOSTWICK}.
This dead layer  carries a large 
density of localised states, and  
charge 
transfer  \cite{ChargeTransfer10,Sasha1,Sasha2,Rozhko,Sasha3, FalkoPinning} between graphene and these surface states dopes graphene.  At  a strong magnetic field, such charge transfer pins graphene doping at integer filling 
factors\cite{FalkoPinning,Sasha3},  leading to the anomalously wide 
QHE plateux,
in particular, at filling factors $\nu = \pm 2 $. 
This makes G/SiC  a promising material platform for the realisation of metrological  resistance 
standard  based on the QHE phenomenon\cite{Sasha1,Sasha2,Rozhko,Sasha3,FalkoPinning,TransferredGrapheneQHE,
CWDGrapheneQHE,LongPlateau16,CWDTransferred, QHEEpitaxialLowField}. 
For practical applications of G/SiC in resistance metrology, which requires achieving robust QHE plateux at moderate magnetic fields, 
 top gating is used  to reduce graphene 
doping. This should be contrasted  to graphene exfoliated onto an insulating substrate  in a field-effect transistor (GraFET), where gates are used to dope otherwise neutral graphene.
Below, we show that these features of G/SiC, as well as  an efficient electrostatic screening produced by  charge transfer between graphene and surface states on SiC 
suppress the formation of  compressible stripes near graphene edge.

\begin{figure}
\begin{center}
\includegraphics[scale=0.77]{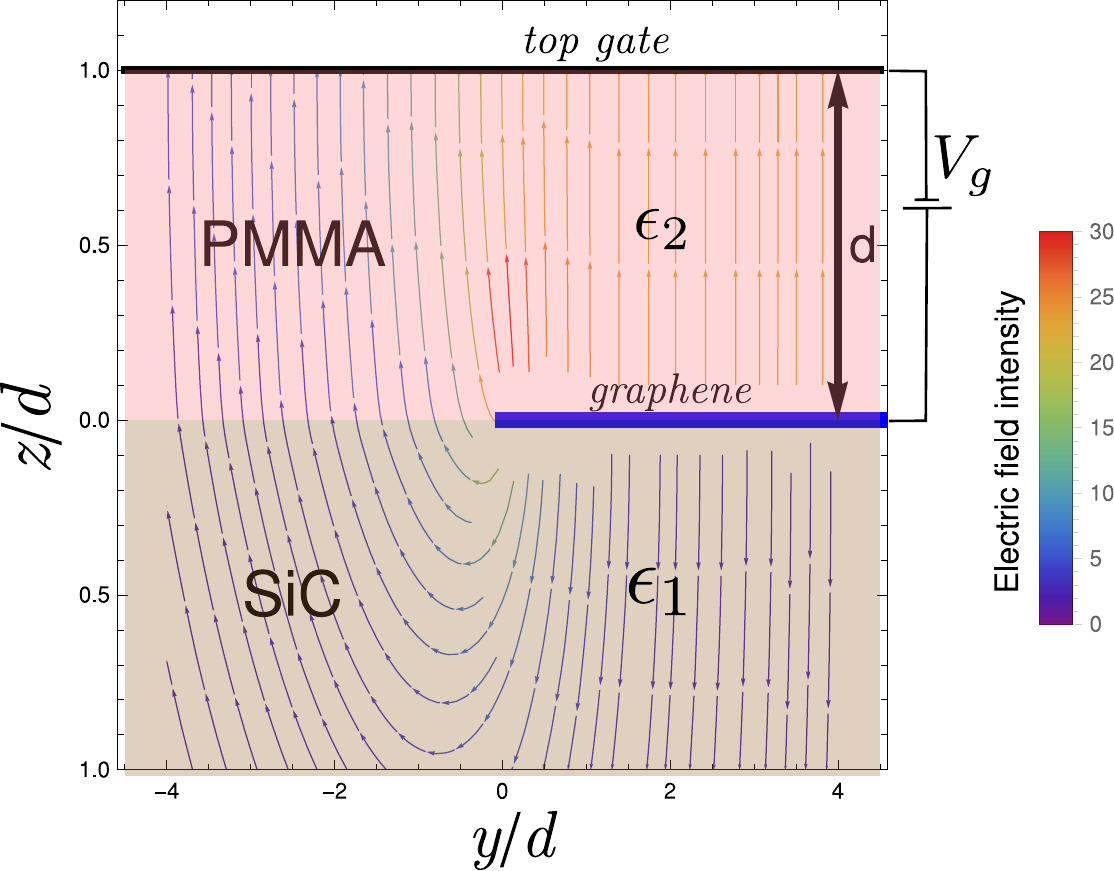}
 \end{center}
\vspace{-0.4cm}
\caption{\label{fig:GatingIllustration} 
Edge of  gated epitaxial graphene on SiC (G/SiC). Electric field lines are plotted with color indicating the field strength.
 The parameters used in numerical analysis  are:    $\eps_1 = 9.7$,  $ \eps_2 = 3.5 $,  $d = 20 \div 400 \nm$.
}
\vspace{-0.5 cm}
\end{figure}

Here, we consider G/SiC  with a top gate located at a distance $d$ above graphene and extended beyond 
the edge of graphene flake, as shown in Fig. \ref{fig:GatingIllustration}.  We assume that, 
far away from the edge ($y \to \infty$), graphene is tuned to the most robust $\nu = \pm 2 $ QHE plateau states.  
In Section \ref{sec:electrostatics}, we model the  electrostatic environment of such gated G/SiC devices. 
This gives an  input for analysing  the self-consistent potential near the edge and for finding the electronic spectrum in quantizing magnetic fields.  In  Section \ref{sec:EpitaxialSpectrum},  
we show that, when a significant potential inhomogeneity builds up, extra pairs of counter-propagating edge channels start to appear  at the edge, whereas the inner edge state may give rise  to a  narrow compressible stripe. 
In Section \ref{sec:StripsStandard}, we compare the edge states in G/SiC with edge states in GraFETs, concluding that in G/SiC formation of compressible stripes is strongly suppressed by
the features of charge transfer in this system.

\section{Edge potential profile for gated G/SiC} \label{sec:electrostatics} 
In this section we analyse the electrostatics of graphene's edge in typical devices  used in graphene QHE metrology \cite{Sasha1,Sasha2,Rozhko,Sasha3}, Fig. \ref{fig:GatingIllustration}, where epitaxial G/SiC  is  coated with PMMA  
and top-gated using a metallic electrode.
It has been noticed \cite{EpitaxialGrowth, ReviewBOSTWICK, ChemicalGating}  that, due to the work function difference between graphene and  SiC surface states,  
epitaxial graphene grown on Si-terminated face of SiC (G/SiC)  is  significantly electron-doped by the charge transfer\cite{ChargeTransfer10,Sasha1,Sasha2,Rozhko,Sasha3, FalkoPinning} from the dead layer of carbons on SiC surface.  
Based on parameters quoted in Refs.[\onlinecite{Sasha3}] and [\onlinecite{LongPlateau16}],  we use $\gamma  \approx 2 \nm^{-2} \eV^{-1}$ for the density of localized surface states on SiC,  
  which appear in the immediate vicinity, $d_1 \approx 0.3 \nm \ll d$, of graphene. 
To reduce the electron doping of graphene and achieve QHE filling factor $\nu =\pm 2$ for  magnetic fields range, $B \sim   5 - 15 $ Tesla, one employs electrostatic gating. 
 Due to the presence of donors just underneath graphene, the compressibility of electrons in this system
never falls below the value determined by $\gamma$, even in places where the Fermi level, $\mu$, falls between Landau levels in graphene.
Therefore, for so large $\gamma$  that  
$$ \frac{\gamma e^2 d}{2 \pi \eps_0 (\eps_1+\eps_2)} \equiv \tilde \gamma \gg 1, 
$$ 
{the quantum 
capacitance of  graphene together with a surface dead layer is much larger than the 
geometric capacitance, resulting in a metallic behaviour of this system}.
For G/SiC,  $\tilde\gamma$ falls in the range $10<\tilde \gamma <200$ for the device parameters listed in Fig. \ref{fig:GatingIllustration}, allowing
us to consider graphene electrons and SiC surface donors, with a total charge density, 
\beq  \label{TotalCharge}
\rho(y)=-e n_G(y) + e n_{D^+},
\eeq
as an almost perfectly screening charge system on the half-plane, $y>0$, $z=0$.
Imposing the condition  $\varphi(y,d) = 0$ at the metallic top gate,  one finds\cite{Efetov}
\beqa  \label{phi12}
\varphi(y,z<0) &=&  \frac{1}{ 2 \pi \eps_0} \int\limits_0^\infty \left[  - \frac{2 \rho(y')}{\eps_1 + \eps_2} \ln |{\bf R}| \right. \hspace{2cm}  \\ 
  &&+\left.\sum_{n=1}^\infty \frac{4 \eps_2 \rho(y') \xi^n}{\eps_1^2-\eps_2^2} \ln |{\bf R} - 
2 n {\bf d}|   \right] dy' , \nn \\
  \varphi(y,z>0) &=& \int\limits_0^\infty \sum_{n=0}^\infty \frac{2  \rho(y') \xi^n}{2 \pi \eps_0 (\eps_1+\eps_2)} \ln\frac{|{\bf R} - 2 (n+1) {\bf d}|} {|{\bf R} +2 n {\bf d} |}  dy'. \nn
 \eeqa    
 Here  ${\bf R} = \left ( 0, y-y',z \right) $, ${\bf d} = (0,0 ,d)$ and  $\xi = \frac{\eps_1 - \eps_2}{\eps_1+ \eps_2}$.
Also,  
\beqa \label{KernelK}
 \varphi(y,0) &=&  \int_0^\infty K(y,y') \rho(y')  dy' \equiv \varphi(y)  ,   \\ 
 \nn \ K(y,y') &=& \sum_{n=0}^\infty \frac{\xi^n}{2 \pi \eps_0 (\eps_1+\eps_2)} \ln\frac{(y'-y)^2 + 
4 (n+1)^2 d^2}{ 
(y'-y)^2 + (2 n d)^2}.   
\eeqa
The electric field $\mathcal{E}_z$ just above graphene ($z=0+$) is  
\beqa
  \mathcal{E}_z(y)& \equiv &  - \d_z \varphi(y,z)|_{z=0+} =  \int_0^\infty K_\mathcal{E}(y,y') \rho(y') dy', 
 \nn  \\ \nn K_{\mathcal{E}}(y,y') &=& \sum_{n=0}^\infty \frac{ \xi^n\, d^2}{2 \pi \eps_0 (\eps_1+\eps_2)}
  \left[ \frac{4 n}{4 (n d)^2+\left(y-y'\right)^2}\right. \\ 
  &&\hspace{0.8cm}+\left. \nn \frac{4 (n+1)}{4 (n+1)^2 d^2 +\left(y-y'\right)^2} \right]. 
\eeqa
The condition for electrochemical equilibrium for the composite system of graphene and donors is, 
\beq \label{PhiChemical}
 \varphi(y) + \frac{\rho(y) - \rho(\infty)}{e^2 \gamma} = - V_g,  
\eeq
and this leads to an inverse problem to find the charge density for a given voltage $V_g$. 

Analytical solutions of Eqs. (\ref{KernelK}) and (\ref{PhiChemical}) are known for $\gamma \to \infty$ in several asymptotic limits: 
\begin{enumerate}[label=(\alph*), wide, labelwidth=!, labelindent=0pt]
\item For  $\eps_1 \to \infty$,  the substrate acts as a metal along the whole plane and the exact solution 
 is $\rho = const$, corresponding to infinite plane  capacitor.
\item For finite $\eps_{1,2}$, the solution to  \Eqs{KernelK}{PhiChemical} has  a singular behavior\cite{TextBook, KhaetskiiFalko, Efetov} near the edge of 
graphene, 
$$
\rho(y) \sim  \mathcal{E}_z(y)  \sim y^{-1/2},  \text{ for }  \tilde\gamma^{-2}  < \frac{y}{d} <  1,
$$
while a finite density of states  regularizes the divergence at  $y < d/\tilde\gamma^2 $, and the presence of a metallic top-gate is responsible for a stronger decay at $y \gg d$.   
\item  When  $\eps_1 = \eps_2=\eps$ ($\xi = 0$ in \Eq{phi12}), 
the 
problem is reduced to finding the charge distribution near the ends of a plane capacitor \cite{TextBook,Lavrentiev}. 
A holomorphic mapping of 
infinite strip, $(-V_g \leq \varphi  \leq V_g,  u)$,  to the complex $y-z$ plane with two cuts, $(y>0, \,z=0)$ and $(y>0, \, z=2 d)$, produces a solution for potential $\varphi$, 
\beq \nn
  y + i (d-z) = - \frac{d}{\pi} \left[ e^{\frac{\pi}{V_g} (i \varphi + u)} + \frac{\pi}{V} (i \varphi + u) + 1 \right].
\eeq
Exactly on graphene 
($z=0$, $\varphi = -V_g$), the auxiliary variable $u$ is related to $y$ via
$$\frac{\pi y}{d} =  e^{\frac{\pi u}{V_g}}-\frac{\pi u}{V_g} -1 $$ 
with $u< 0$ ($u>0$) corresponding to points just above (below)  graphene (i.e. $z=\pm 0$). 
Electric field near graphene is $\mathcal{E}_z = \frac{V_g}{d} \frac{1}{1 - e^{\frac{\pi u}{V_g}}}$ and the charge density at $z=0$ is given by
$$\rho(y) =\frac{\mathcal{E}_z(y,+0) - \mathcal{E}_z(y,-0)}{\eps\eps_0} \underset{y \gg d}{\approx} \frac{V_g}{\eps_0 \eps d} \left( 1 + \frac{d}{\pi y}  \right ). $$
 \item  For $\eps_1 \ll \eps_2$, one can find \cite{KhaetskiiFalko} for the electrostatic problem:
 \beq \nn
\rho(y) = \frac{ - \eps_0 \eps_1 V_g }{d \sqrt{1-e^{-\frac{\pi y}{d}}}}. 
\eeq 
\end{enumerate}

Aiming at modelling the devices used in the experiments reported in  Refs.[\onlinecite{Sasha2, Sasha3, Rozhko}], we 
solve \Eqs{KernelK}{PhiChemical} numerically for $\tilde\gamma \gg 1$. Then, knowing the form of all of the above-listed asymptotics, we find the  interpolation formula,
 \beqa
\rho(y) &=& \frac{-3.5 V_g \eps_0}{d} \left( \frac{1}{\frac{1}{2 \tilde \gamma} + \sqrt{1-e^{-\frac{\pi y}{2 d} } }} +  \frac{0.5}{\frac{1}{ \tilde \gamma} + \sqrt{\frac{y}{d}}}  \right), \nn\\
  \mathcal{E}_z(y) &=&  \frac{-V_g}{d} \left(1+ \frac{0.22 e^{-\frac{2.5 y}{d}}}{\frac{1}{ \tilde \gamma}+\sqrt{\frac{y}{d}}} \right),
 \label{rhoEq} 
 \eeqa
which work with 1\% accuracy for the obtained numerical solution. 


\section{Compressibility of QHE edge states in G/SiC} \label{sec:EpitaxialSpectrum}
Having found the total charge density of graphene electrons and SiC surface donors, we find how the  total charge, $\rho$, is divided between graphene, $-e n_G$, and SiC donor states, $e n_D$.
We relate the electrostatic potential  for  surface states in SiC, $U_{SiC}$,   to electrostatic potential {energy $U$ of electrons} in graphene 
as:  
\beq \label{EqUSiC}
  U_{SiC}(y) = U(y) - e d_1 \left[\eps_2 \mathcal{E}_z(y) + \frac{e n_G(y)}{\eps_0}\right].
\eeq
Here 
$n_G(y)$ is the local 
electron density in graphene at point $y$ and 
the density of donors on SiC surface is \cite{ChargeTransfer10,Sasha1, FalkoPinning}: 
\beq \label{Eqmu}
  n_{D^+} = \gamma \left[ A +  U_{SiC}(y) - \mu\right],
\eeq
where $A \approx 0.2 \,\eV$ is a work function difference between charge-neutral SiC surface and undoped graphene.
Electrochemical equilibrium conditions require that electrons in localized surface states on SiC have the 
same electrochemical potential, $\mu$, as graphene, which we count from the Dirac point in graphene far away from the edge,
so that
\beq  \label{ChargeTransfer2plate}
  \ba{c} \frac{n_G(y)}{\gamma_{\rm eff}}  = 
   A - \mu + U(y) - e \eps_2 \mathcal{E}_z(y) d_1  - \frac{\rho(y)}{e \gamma},\\
   \frac{1}{\gamma_{\rm eff}} \equiv  \frac{1}{\gamma} + \frac{d_1 e^2}{\eps_0}.
  \ea
\eeq

\begin{figure}
 \includegraphics[scale=0.492]{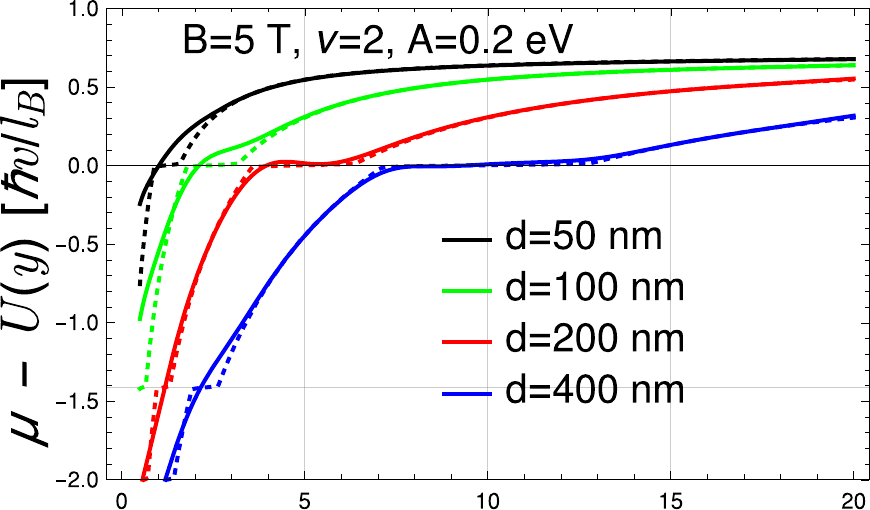} 
\includegraphics[scale=0.487]{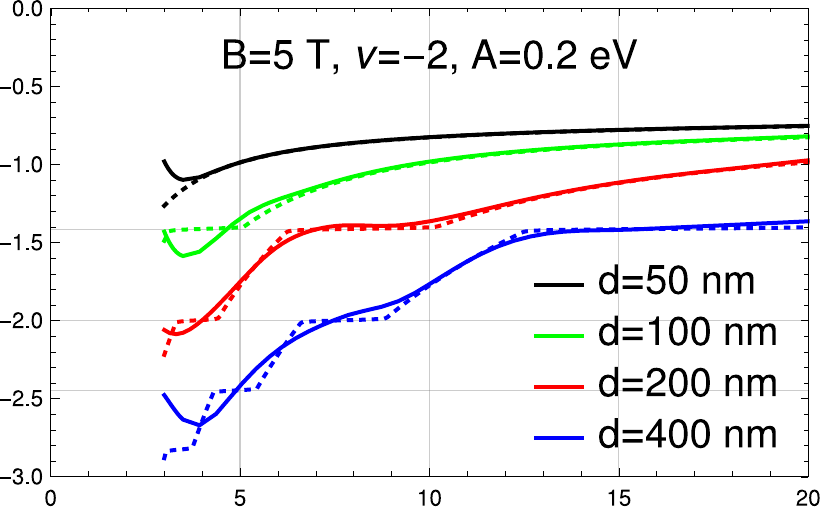}\\ 
\hspace{-0.74mm}
 \includegraphics[scale=0.4915]{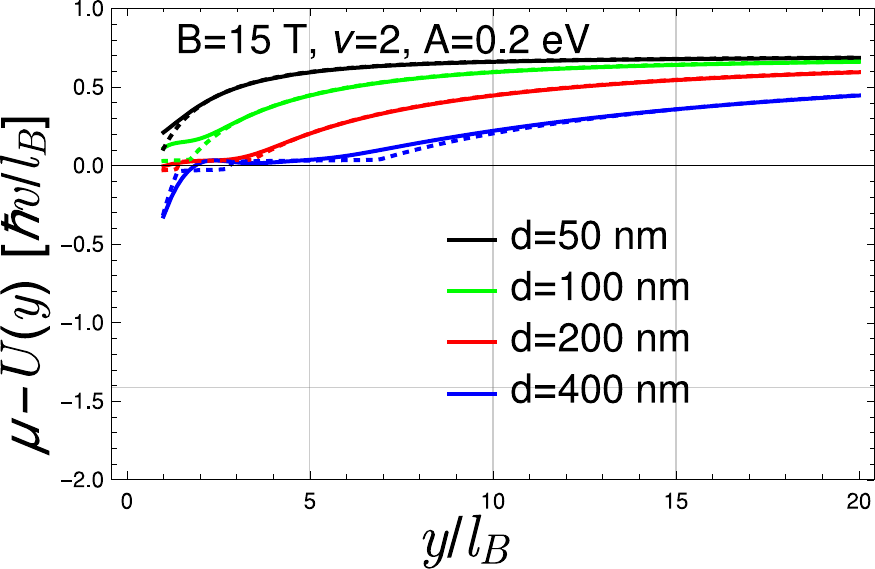} 
\includegraphics[scale=0.488]{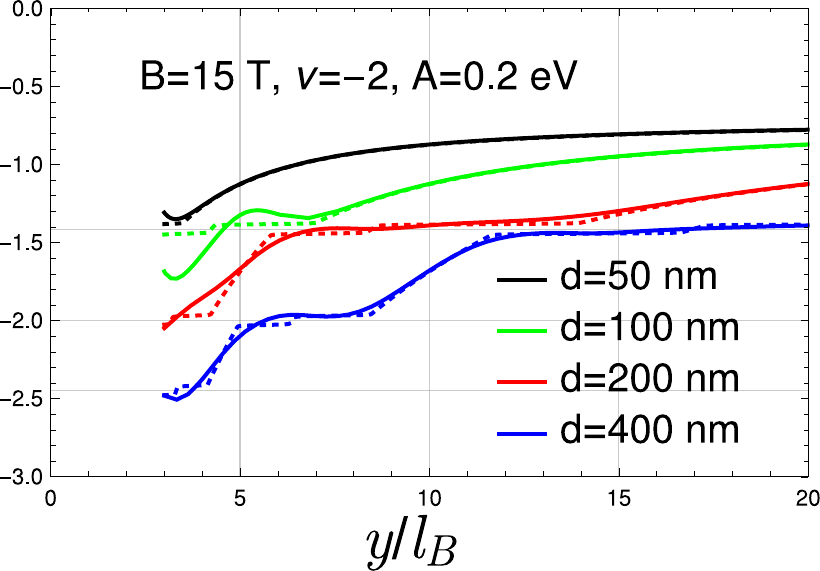}
\vspace{-0.1cm}
\caption{\label{fig:Stripes} 
Self-consistent electrostatic potential induced  in G/SiC   for $B=5$ T and $B=15$ T,  bulk filling factor $\nu=\pm 2$ (left/right columns)  $d=50, 100, 200, 400 \nm$ (top to bottom).
Fermi level is chosen as $\mu = \pm \frac{1}{\sqrt{2}} \frac{ \hbar v }{l_B} $, corresponding to the middle of the gap for $\nu = \pm 2 $. 
{Dashed line corresponds to quasiclassical calculation, \eq{FillingClassical1}; solid lines are self-consistent calculation, \eq{StripsSiC}. } 
For $B=15$ T, we illustrate the effect of interaction-induced level splitting by $6 \, \rm m\eV$.}
\end{figure}

In general, the relation between $U(y)$ and $n_G(y)$ is  non-local 
[see \eq{nGEquation} below]. However, in the case that  potential $U(y)$  varies slowly 
at the length scale of magnetic length, $l_B = \sqrt{\frac{\hbar}{eB}}$,  ($\nabla U \ll \frac{\hbar v}{l_B^2} $),  one may 
approximate the local energy of Landau levels (LL) in graphene as
\beq
E_n + U(y) = \frac{\hbar v}{l_B} \sign(n) \sqrt{2 |n|} + U(y).
\eeq
The local density of electrons can be related to the local filling factor, $\nu_n(y)$,
determined by the number of filled LLs and spin/valley degeneracy,
\beqa 
  \label{FillingClassical1} 
  n_G(y) \approx \frac{e B}{h} \nu(y).
   \eeqa
Solving \eq{ChargeTransfer2plate} with $n_G$ given above leads to results  shown by dotted lines in Fig. \ref{fig:Stripes}.
The resulting potential on graphene have a number of horizontal intervals, which would correspond to compressible stripes,
if the conditions for validity of  quasiclassical approximations 
are satisfied. 

 When the width of a stripe is comparable to magnetic length,  $l_B$, 
 we must account for a
 finite extent of electron's wave functions in a magnetic field, therefore,  going beyond the quasi-classical approximation used in \eq{FillingClassical1}.  
Here, we use Landau gauge and  parameterise states  with  momentum $p$ along the edge,  related to the distance of electronic 
wave function from the edge, $y_0 =  - l_B^2\, p $.  
When $y_0 \gtrsim 2\, l_B$, wave functions of LLs have a Gaussian form:
\beqa
\nn
|\psi_{0,y_0}(y)|^2 &=&\frac{1}{ \sqrt{\pi} l_B} e^{-\delta^2} , \ \ \  \delta \equiv \frac{y-y_0}{l_B};\\
|\psi_{n\neq0,y_0}(y)|^2& =&  \frac{ H^2_{|n|}(\delta) + 2 |n| H^2_{|n|-1}(\delta)}{2^{|n|+1} |n|! \sqrt{\pi} l_B} e^{-\delta^2},
\nn
\eeqa
 where $H_n(x)$ are Hermite polynomials.
While for a smooth  potential $U(y)$, 
$E_{n,y_0}\approx E_n+U(y_0)$, but, due to potential having cusp-like features  near the ends of compressible stripes (see Fig. \ref{fig:Stripes}), we use a more precise expression
for the LL energy and electron density in graphene, 
\beqa 
E_{n,y_0} &\approx& \int |\psi_{n,y_0}(y) |^2 U(y) dy,  \\
 \label{nGEquation}
 n_G(y) &=&\frac{1}{2 \pi l_B^2}  \sum_{n=-N}^N \int |\psi_{n,y_0}(y)|^2 \tilde\nu_n(y_0)  dy_0, \nn \\
 \tilde\nu_n(y_0) &\equiv&  2\, {\rm Erf}\frac{\int |\psi_{n,y_0}(y)|^2 [\mu - U(y) - E_n] dy }{\Delta}, \nn
 \eeqa
 where we introduced a small Gaussian LL broadening, $\Delta$,  and a LL cut-off, $N$, for the convergence of our numerical procedure.
Note that the result is independent of  $N$  when $|\mu - U(y)| <= \frac{\hbar v}{l_B}\sqrt{2 N}$. 
 Together with \eq{ChargeTransfer2plate}, \eq{nGEquation}  results in a non-linear integral equation for $U(y)$,
 \begin{widetext}
\beq \label{StripsSiC}
\mu - U(y) = - \frac{1}{\gamma_{\rm eff}} \frac{1}{\pi l_B^2}  \int dy_0 \left[ \sum_{n=-N}^N  |\psi_{n, y_0}(y)|^2 
{\rm Erf}\frac{\int |\psi_{n,y_0}(y)|^2 (\mu -U(y)-E_n) dy }{\Delta} \right]    +A - \frac{\rho(y)}{e \gamma} - e \eps_2 \mathcal{E}_z(y) d_1.
\eeq
\end{widetext}
 The latter  equation can be solved iteratively\footnote{The function $U(y)$ is interpolated with $\sim 50 $ points and iterations are done as $U_{new} = \alpha U(U_{old}) + (1- \alpha) U_{old}$ with 
 a factor $\alpha \leq 1 $ small enough to reach convergence.
 At small $\Delta$ the non-linearity gets strong and we need to choose rather small values $\alpha \sim 0.001$ to achieve convergence.}.
 The results of numerical solution of \eq{StripsSiC} 
are shown in Fig. \ref{fig:Stripes},  
where flat intervals of potential (precursors of compressible stripes)  are formed: 
those are about $\sim  l_B$ (or $\sim 2\, l_B$ for $\nu = -2$) narrower  than what was expected 
from the quasi-classical estimations.  Note that the non-locality in \eq{StripsSiC}  arizes due to a finite ($|y-y_0|\sim l_B$) extent of the wave function, rather than long-range Coulomb interactions (as 
 it would be in a  2DEG in GaAs/AlGaAs \cite{Gerhards88, Shklovskii92, Chalker93, Fogler94, GerhardsCurrents}   or in a GraFET, Section \ref{sec:StripsStandard}).
 {We note a significant particle-hole ($\nu = \pm 2$) asymmetry of the potential profile that is caused by a significant initial n-doping of G/SiC: the gate potential has the same sign for both $\nu = +2 $ and $\nu = -2$
 but it is stronger for $\nu = -2$.}
\begin{figure*}
 \includegraphics[scale=0.34]{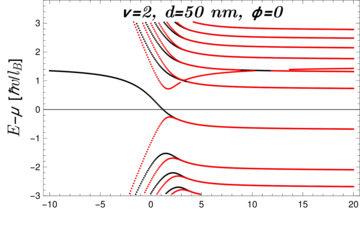}   \includegraphics[scale=0.33]{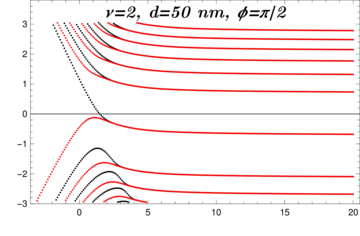}  \includegraphics[scale=0.33]{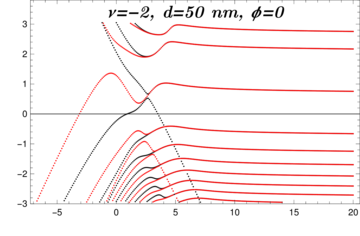}   \includegraphics[scale=0.33]{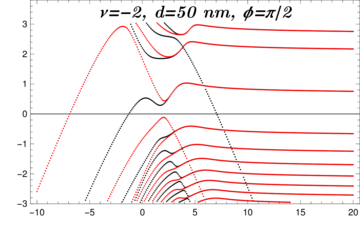} \\
    \includegraphics[scale=0.34]{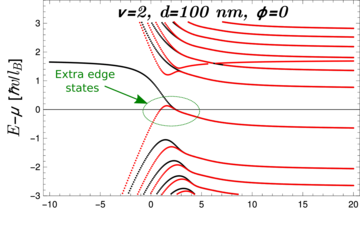}   \includegraphics[scale=0.33]{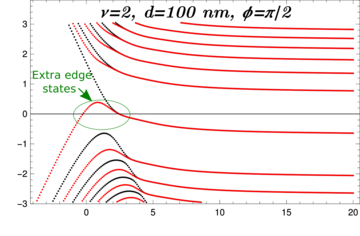}  \includegraphics[scale=0.33]{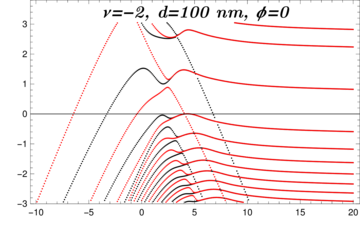}   \includegraphics[scale=0.33]{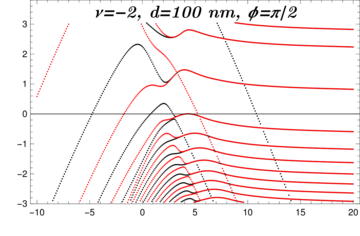}\\
  \includegraphics[scale=0.34]{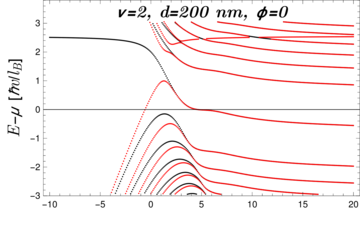}   \includegraphics[scale=0.33]{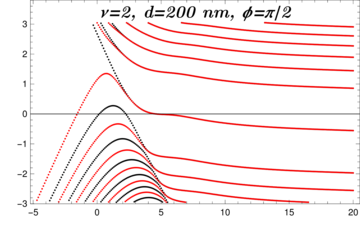}  \includegraphics[scale=0.33]{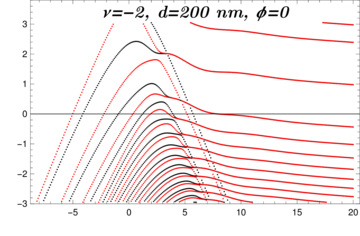}   \includegraphics[scale=0.33]{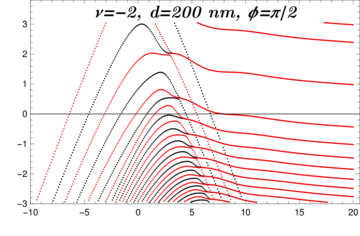}\\
   \includegraphics[scale=0.34]{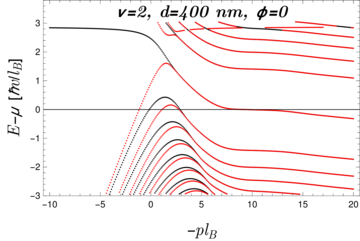}   \includegraphics[scale=0.33]{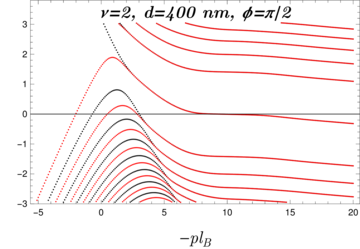}  \includegraphics[scale=0.33]{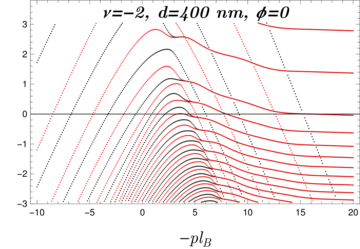}   \includegraphics[scale=0.33]{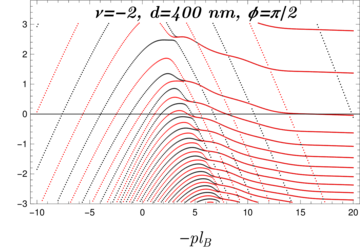}
\vspace{-0.3cm}
 \caption{\label{fig:spectrum} 
Electronic spectrum in magnetic field $B=5$ T calculated for two examples of boundary conditions: $\phi=0$  and $\phi = \pi/2 $  and for filling factors $\nu = \pm 2$. 
Edge states correspond to levels crossing the Fermi-level,
compressible stripes correspond to flat intervals of dispersion at $E  = \mu$. Black/red color corresponds  to different $\pm K$ valleys. Extra branches for $\nu=-2$ are the evanescent waves, localized in the area of large 
potential near the edge (similar to ``whispering gallery'' modes in Ref. \onlinecite{WhisperingGallery0,WhisperingGallery}).
}
\end{figure*}

Also, at distances $\sim l_B$ near the edge, the electronic  wave functions do not have Gaussian shape, 
but this 
can be incorporated in the change of boundary conditions for Dirac electrons, as discussed below\footnote{
The boundary conditions can be used to incorporate the sample-dependent variations of potential near the edge of graphene, 
which arise due to chemical and electrostatic modification of the edge. 
To consider the effect of a large potential $U(y)$ near the edge, 
let us start solving the Dirac equation from a point $y_0>0$,  hence,  we need a new boundary condition, $\phi(y_0)$. 
We find that \Eq{DiracGen}
leads to 
$$
 \frac{d}{d y} \phi(y) = 2 \frac{U(y)-E }{\hbar v}  + 2 \(p+\frac{y}{l_B^2}\) \pm \sin \phi(y),  
$$
producing a flow of boundary conditions for valleys $\pm K$ (we do not loose solutions in this procedure if the second, nonlinear,  term 
can be neglected relative to the first term).
This flow depends  on valley, $p$  and $E$, producing a more general class of $E$-- and $p$-- dependent boundary
conditions.  Considering the flow in the region of large $|U|$  ($|E| \ll |U(y)| $ for $y<y_0$), and concentrating on  low $|p|$, 
leads to an approximate equation 
$\frac{d}{d y} \phi(y) \approx  2 \frac{U(y)}{\hbar v} $.  
It means that when $U(y)$ is positive (negative) near the edge, the 
effective boundary parameter  
increases (decreases) as we move away  from the edge.
As a result, the sample-dependent details of the potential near the edge can be absorbed into a renormalization of $\phi$. 
For example, the standard zigzag boundary conditions,  $\phi(0)=0$, behave similarly to $\phi \approx \pi/4$ in  G/SiC at $\nu=2$.
Note that at large and wide potential near the edge (as happens for $\nu=-2$ in our discussion), moving $y_0$ to the safe region of low potential will involve  $\phi$  
making several $2 \pi$ turns,   resulting in the loss of several solutions, corresponding to extra evanescent modes\cite{WhisperingGallery0,WhisperingGallery}}.
Generic boundary conditions \cite{Akhmerov1, FalkoBoundary, SSFalko}
for graphene electrons near the edge are: 
\beq 
\ba{c}
  v  \bm{\sigma} \cdot (-i \hbar {\bf \nabla} + e {\bf A}) \Psi = (E - U(y)) \Psi \; ;  \\
  \left[ 1 - ({\bm m} \cdot \bm{\tau}) \otimes ( {\bm n} \cdot \bm{\sigma} )]  \Psi \right|_{y=0}  = 0; \\
  {\bm n} =  \hat{\bf n}_z\, \cos \phi +  [{\hat{\bf n}_z} \times {\bf n_\perp}]\, \sin \phi .
 \ea
 \label{DiracGen} 
\eeq
Here,  $\sigma_i$ and $\tau_i$ are Pauli matrices acting separately on sublattice ($A,B$) and valley ($\pm K$) components of a 4-spinor,  
$\Psi^T = (\Psi_{KA}, \Psi_{KB}, \Psi_{-KB}, -\Psi_{-KA})$, describing
the electron amplitudes on sublattices $A$ and $B$ in the valleys $\pm K$. Coordinate axes here are the same as in electrostatics analysis: electrons move freely in half-plane: $-\infty<x<\infty, \,y>0,\, z=0$
with the straight edge $(x,y=0,z=0)$. 
Generic boundary conditions \cite{Akhmerov1, FalkoBoundary, SSFalko}  in \eq{DiracGen} are parameterized by two unit vectors, $\bm m$ 
and 
${\bm n}  \perp {\bf n_\perp}$, where $\bf{n}_{\perp}$ is  normal to
the edge and lays within the  2D plane of graphene.  Both $\bm m$ and $\bm n$ depend on the microscopic features of the edge in a particular sample. 
A rotation of multi-spinor $\Psi$ in the valley space can be used\cite{SSFalko} to set $\bm m = \hat{\bf n}_z$,  so that angle $\phi$ (corresponding to the direction of  vector $\bf{n}_{\perp}$) is the only
relevant boundary parameter  ($\phi \in [0,\pi]$ and  $\phi \to \phi + \pi$ is obtained by swapping the valleys). 


For calculating edge states, we use Landau gauge for vector potential, ${\bf A} = (B y, 0)$, and characterise states by wave-number $p$ along the edge,  $\Psi(x,y) = e^{i p x} \psi(y) $.
Typical  dispersions $E(p)$  are shown in Fig. \ref{fig:spectrum}.  
These  spectra  are  valley-degenerate  away from the edge, while at distances $y\lesssim 2\, l_B$ ($-p\, l_B < 2 $) the valley degeneracy is broken. 
When the top gate is close to graphene, e.g.  $d \lesssim 50 \nm $, the spectrum is qualitatively similar to the case of zero potential\cite{SSFalko,EdgeStates1, EdgeStates2, Abanin2007, EdgeStates3, AbaninEdgeWithScreening}, 
though with some renormalization of $\phi$, caused by 
potential variation at short  distances ($y \lesssim l_B$) near the edge:  we find\cite{Note3} that $\phi$ is effectively increased (decreased) by a positive (negative) peak of potential near the edge of graphene.   
For such close gates,  there is only one edge channel (per spin) with chirality  prescribed by the bulk filling factor ($\nu=2$, or $\nu = -2$). 
For $d\gtrsim 100 \nm$, the first pair of counter-propagating 
edge channels starts to appear, as pointed in Fig. \ref{fig:spectrum} by an arrow. 
We  find that for $d \gtrsim  20 \, l_B$ ($d>200 \nm$ in Fig. \ref{fig:spectrum}) the edge channel  starts to develop a narrow  valley-degenerate  compressible stripe.

Besides the above-listed features, common for $\nu=2$ and $\nu=-2$,  there are the following notable differences between those two filling factors.
First of all, extra pairs of edge states  and compressible stripes correspond to the 0-th LL  for $\nu = 2$ and to the ``$n=-1$''  LL  for $\nu = -2$. 
Then, apart from the expected continuation of the bulk Landau levels,  we observe extra branches of evanescent edge modes that  generalize the
zigzag edge modes\cite{EdgeStates1, EdgeStates2, EdgeStates3, Abanin2007, AbaninEdgeWithScreening} to generic boundary conditions.  
For $\nu=2$, there is only one such branch in each of the valleys, with its dispersion depending on the renormalized value of $\phi$.  
It approaches the Fermi level at low values of $p\, l_B$, resulting in  strong  mixing with LL branches and multiple avoided crossings.  
In G/SiC reaching $\nu=-2$ requires larger top-gate voltage, making the effects of external potential stronger. This is reflected in more evanescent modes, that cross  Fermi level 
at larger values of $|p\, l_B|$.  Note that these  edge modes are present even in a zero magnetic field and may be explained by full internal reflection from the potential wall\cite{WhisperingGallery0, WhisperingGallery}.     

\begin{figure*}
\begin{center}
  \includegraphics[scale=0.825]{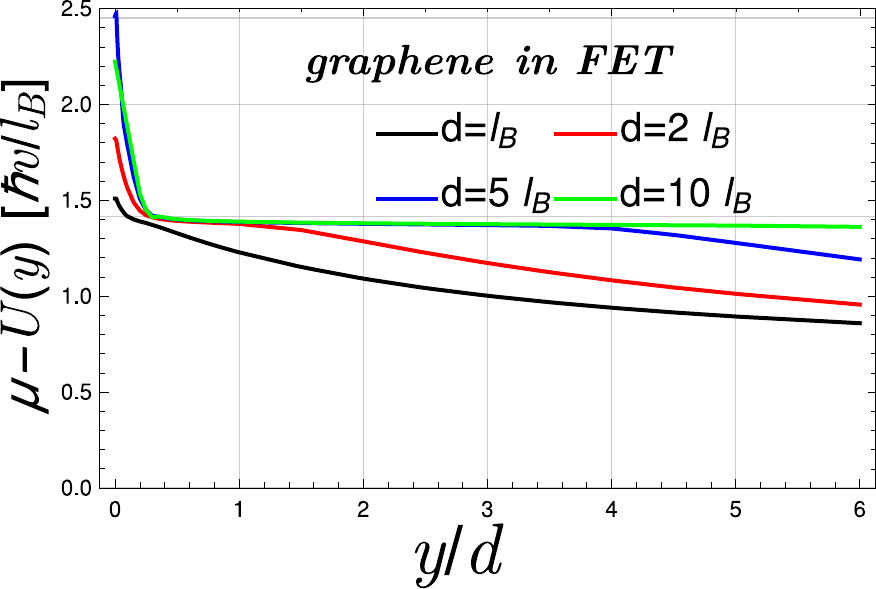} 
  \includegraphics[scale=0.8]{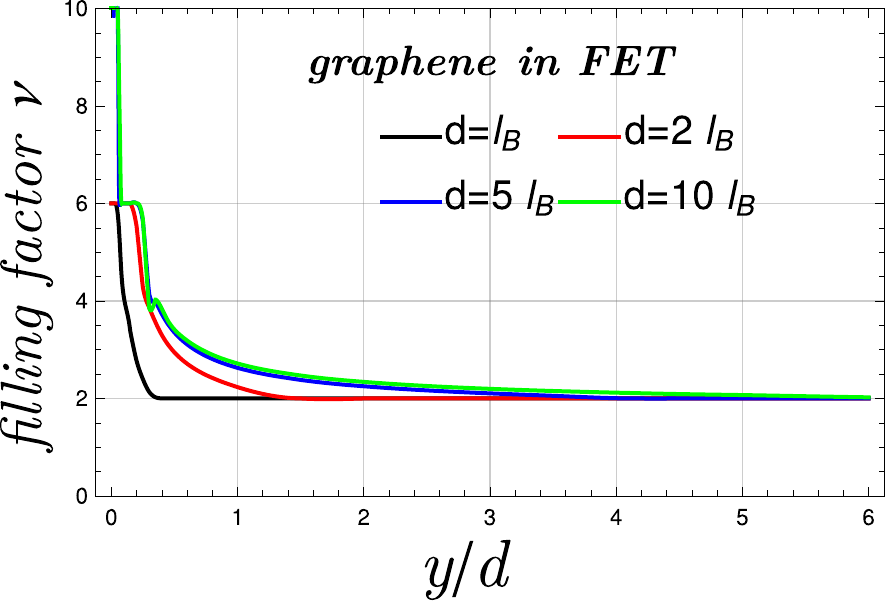} 
\vspace{-0.2cm}
  \caption{\label{fig:SiO2} 
Self-consistent potential (local Fermi energy),  $\mu - U(y)$, and the filling factor, $\nu$, in GraFET. 
Top gate voltage $V_g$ is tuned to achieve a midgap chemical potential in the center of a device, chosen to have a width $50\, d$.
Long-range Coulomb interaction leads to  wide compressible stripes (corresponding to LL with $n=1$) developing already for $d \gtrsim 2\, l_B$. 
}
\end{center}
\end{figure*}
\begin{figure}
\begin{center}
\includegraphics[scale=0.34]{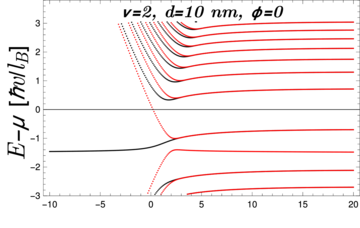}   \includegraphics[scale=0.33]{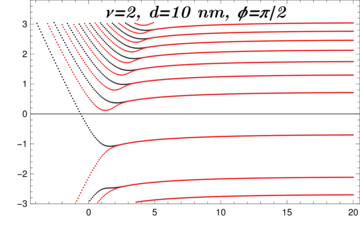}\\
  \includegraphics[scale=0.34]{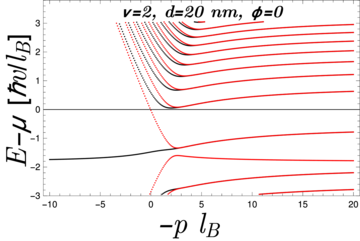}   \includegraphics[scale=0.33]{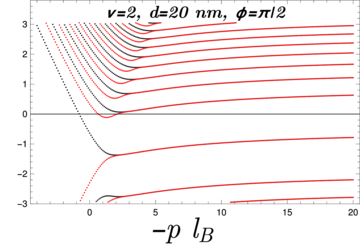}\\
   \includegraphics[scale=0.5]{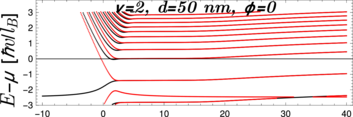}  \\  \includegraphics[scale=0.5]{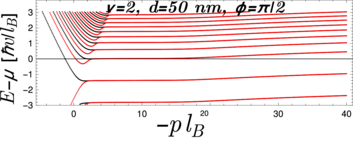}
  \end{center}
  \vspace{-0.4cm}
\caption{\label{fig:spectrumFET}
Edge state dispersion in GraFET with boundary condition parameters $\phi=0$ and $\phi=\pi/2$, for $B = 5  \, \rm T$. Black/red color corresponds  to different $\pm K$ valleys.
Compressible stripe starts to develop already for $d>2\, l_B \approx 20 \nm$, getting  $\L \sim 20\, l_B$ wide at $d=5\, l_B$ and rapidly growing further with increasing $d$.
For $\phi=0$ note the appearance of an almost flat branch of evanescent edge modes below the Fermi level. 
}
\end{figure}

\section{Compressibility of QHE edge states in $\text{GraFETs}$} \label{sec:StripsStandard}
For comparison, in GraFETs,  graphene doping is provided solely by electrons transferred from the gate.
In this case, we solve
a self-consistent nonlinear integral equation\cite{Gerhards88, Shklovskii92, Chalker93, Fogler94, GerhardsCurrents}, desribing potential near GraFET edge,
\beq \label{KernelK3}
   U(y)  =\frac{\hbar v}{l_B} \kappa \int\limits_0^\infty  \frac{dy'}{d} \nu(y')   \sum_{n=0}^\infty \xi^n \ln\frac{(y'-y)^2 + 
4 (n+1)^2 d^2}{ 
(y'-y)^2 + (2 n d)^2}  ,
\eeq
with
$
  \kappa = \frac{e^2}{(2 \pi)^2 \hbar v \eps_0 (\eps_1+\eps_2)}\frac{d}{l_B} \approx 0.05 \frac{d}{l_B} \nn
$ {(we use  $\eps_1 = 9.7$,  $ \eps_2 = 3.5 $)}
and 
$ \nu(y) = \sum_{n=-N}^N  2\, {\rm Erf}\frac{\mu-U(y)-E_n}{\Delta} $. 
The nonlocality in it is produced by the long-range nature of Coulomb potential, so that the ``edge-of-capacitor'' effect extends over longer distances, as compared to G/SiC system.
This supports the use of quasi-classical approximation, $ 2 \pi l_B^2  n_G(y) = \nu(y) $, which is then  justified  by the results, Figs. \ref{fig:SiO2} and \ref{fig:spectrumFET},  
showing compressible stripes with widths $\L \,\gg l_B$ that appear at $y \gg l_B$.    

Choosing the top-gate voltage,  
$$ e V_g =\frac{\hbar v}{l_B} \left[ 2 \kappa \frac{2 \pi (\eps_1 + \eps_2)}{\eps_2} + \frac{1}{\sqrt{2}}\right], $$  
to get  $\nu = 2$ away from the boundary\footnote{
This equation is easily derived from the plain capacitor geometry and the requirement $\mu = \frac{\hbar v}{l_B}  \frac{1}{\sqrt{2}}$,  $\rho =2  e /(2 \pi l_B^2) $  
}, 
we solve \eq{KernelK3}  numerically.   For numerical simulations\cite{Note2}, we set  width of $50\, d$ for the modeled GraFET device  and tune  chemical potential 
to the middle of the gap in the center of the device.  
The results of numerical simulations are shown in Fig. \ref{fig:SiO2}  and the LL spectrum is plotted in Fig. \ref{fig:spectrumFET}.  Results for  $\nu = -2$ can be obtained by reversing 
the sign of energy and changing\cite{SSFalko} $\phi \to -\phi$. 
As compared to the case of G/SiC, the opposite (positive) sign of top gate voltage is needed to dope 
graphene to the  
filling factor $\nu=2$, which leads to
a stronger electron doping near the edge. As a result, for increasing $d$, extra counter-propagating edge channels  correspond to filling of the 1-st LL, in contrast to the 0-th LL 
in case of $\nu=2$ in G/SiC.    
Another difference is that the inner edge channel  reconstructs easier into  a wide compressible stripe (now, for the 1-st LL). 
This starts to happen already at $d \sim 2 l_B$, and the compressible
stripe rapidly grows upon increasing $d$, see Figs. \ref{fig:SiO2} and \ref{fig:spectrumFET}. 
This latter remark can be used to interpret  the recent experiment \cite{EdgeSiO2}, 
where the formation of a wide compressible stripe has prevented the edge metallization contacts from measuring the $\nu = 2$ resistance plateau, despite the $\nu=2$ incompressible state in the bulk of the sample.   

\section{Discussions and conclusions}  \label{sec:Conclusions}
The analysis of the electrostatics of QHE edge states in graphene presented in Sections \ref{sec:electrostatics} - \ref{sec:StripsStandard}
establishes that formation of compressible stripes near the edge is suppressed in G/SiC, as compared to GraFETs and 2DEG in GaAs/AlGaAs heterostructures. 
  Numerical solutions of \Eqs{StripsSiC}{KernelK3} for the potential and \eq{DiracGen} for the spectrum allows us to calculate the width, $\Delta p$, of flat interval in the dispersion at the Fermi level
  (where 
 $E(p)=\mu$). The corresponding states are located at $y \approx -p\, l_B^2$ thus forming a compressible stripe of width  
 $\L = l_B^2 \Delta p$.       To illustrate the result quantatively,  
  we choose the  device parameters that provide $\nu = \pm 2$ QHE plateaux for $B = 5$ Tesla and $B=15$ Tesla, motivated by
 the  implementation of G/SiC  in the primary\cite{Sasha2} (15 Tesla range) and a ``push-button''\cite{Rozhko} (5 Tesla range)  quantum resistance standards 
 by National Physical Laboratory (UK)  and Oxford Instruments PLC.

 \begin{figure}
\begin{center}
  \includegraphics[scale=0.65]{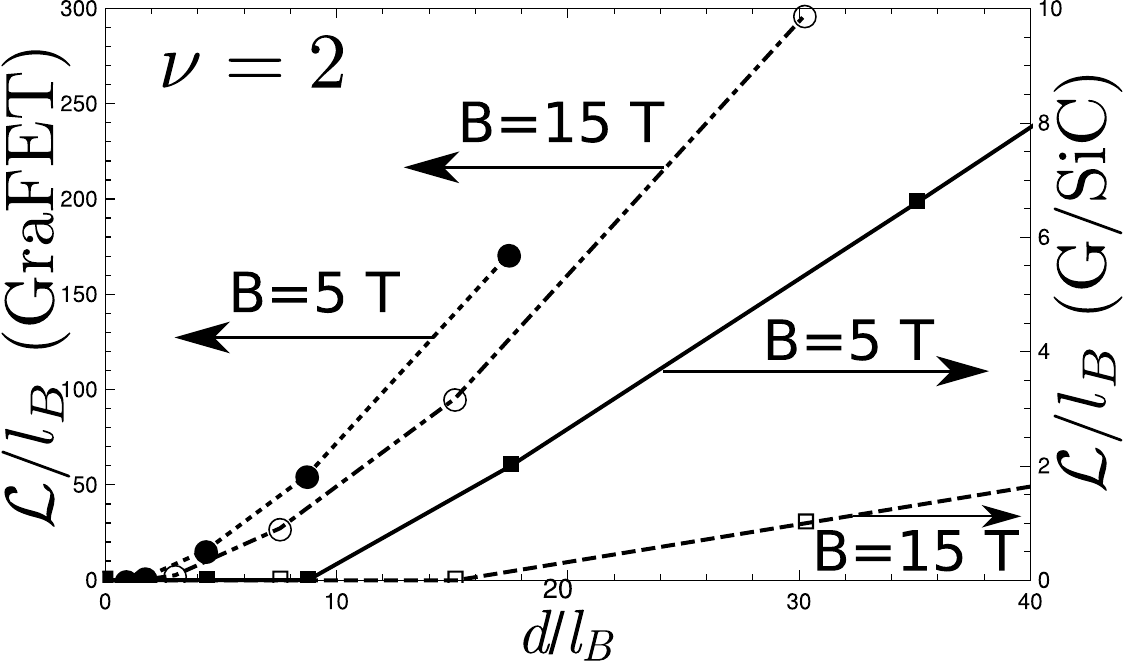}\\ 
  \vspace{0.1cm}
  \includegraphics[scale=0.65]{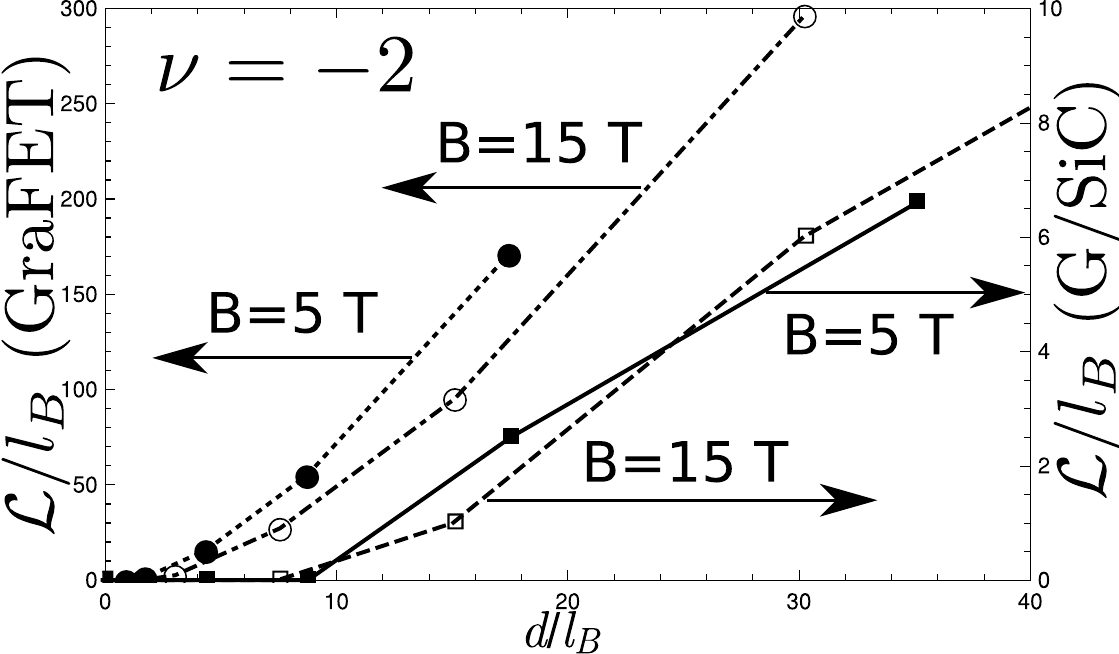} 
  \end{center}
\vspace{-0.7 cm}
\caption{\label{fig:StripeWidth} 
Compressible stripe width, $\L$, near the $\nu = \pm 2$ QHE edge in graphene in G/SiC (squares) and GraFET (circles) 
as a function of a distance to the top gate, $d$, for  $B=5$ T (filled symbols) 
and $B=15$ T (open symbols). Note the difference between r.h.s. and l.h.s. axes, stressing the fact that compressible stripes in GraFETs are much wider than 
in G/SiC.  
}
\end{figure} 
   The results for the compressible stripe length, $\L$,  are gathered in Fig.\ref{fig:StripeWidth} fro both G/SiC and GraFETs.  
They highlight an essential feature of G/SiC determined by  the presence of surface states in the ``dead layer'' of SiC surface, 
which leads to  significant electron doping of graphene (so that  the opposite signs of gate potentials are needed to achieve $\nu = 2 $ in G/SiC and in GraFETs)
and to an efficient electrostatic screening. 
As a result,
compressible stripes in G/SiC are possible only for 
distances $d \gtrsim 20\, l_B$ between graphene and the  gate,  
and even then their widths, $\L$, are order of magnitude smaller than  in  GraFETs,  Fig. \ref{fig:StripeWidth}.
  Semiclassically, one would  estimate\cite{Shklovskii92,KhaetskiiFalko}   $\L$ 
as a distance     
at which the filling factor changes by 4, which would give a linear dependence for G/SiC \footnote{To estimate $\L$ we use \eq{ChargeTransfer2plate}. The dependence $\L \propto d$ follows from the scaling
$\mathcal{E}_z(y ; d) = \tilde{\mathcal{E}}_z (y/d)/d$ and $\rho(y; d) = \tilde \rho (y/d)/d$  in  \Eq{rhoEq}.}, 
$$ 
 \L \approx \left. \frac{4 e B}{h \gamma_{\rm eff} \left|\frac{\d_y \rho(y)}{e \gamma} + \d_y \mathcal{E}_z(y) e \eps_2 d_1  \right|}  \right|_{\hspace{-0.3cm}\ba{cc} & \scriptstyle 
 y=y_c, \\ &\scriptstyle  U(y_c) = \mu \ea } \hspace{-0.2cm} \propto d .    
$$
Note that by the definition in \eq{ChargeTransfer2plate},  $\gamma_{\rm eff} < \eps_0 d_1/e^2$, whereas would it be formally set to $\gamma_{\rm eff} \to \infty$, no compressible stripe could form. 
Also, magnetic field dependence of $\L$  is different from linear, because at fixed filling factor in the 2D bulk, changing magnetic field would be accompanied by changing gate voltage (hence, $\mathcal{E}_z$ and 
$\rho$ are changing).   For example, at large magnetic fields ($B \sim 20$ T), no gate voltage is needed to get $\nu=2$ in the bulk
and no compressible stripes are expected, while for $\nu=-2$ the needed gate volatage increases with magnetic field, which may lead to wider compressible stripe at larger $B$.  
The above quasiclassical estimate agrees with the results shown in Fig. \ref{fig:StripeWidth} 
upon  subtraction of $\sim 2 \, l_B$ for $\nu=2$ and $\sim 4 \, l_B$ for $\nu=-2$, which
accounts for a finite extent, $\sim l_B  $, of electron wave functions in the relevant LLs.
The latter difference is one of the manifestations of the ``electron-hole'' asymmetry of QHE edge states in G/SiC, in contrast to ``e-h'' ($\nu \to - \nu$) symmetry of QHE edge states in GraFETs. 
%
This ``e-h'' asymmetry of G/SiC is determined by that 
reverting graphene doping from n-type ($\nu=2$)
to p-type ($\nu = -2$) 
requires further increasing gate voltage rather than reverting the sign of gate voltage as in the case of GraFET. 
The resulting potential inhomogeneity near the edge is stronger for $\nu=-2$ in G/SiC,   leading to a 
larger number of counter-propagating pairs of evanescent edge 
modes that are present even at zero magnetic field \cite{WhisperingGallery0, WhisperingGallery}.
{Counter-propagating modes lead to dissipative QHE \cite{DissipativeHall07} unless they are not gapped by 1D localization, induced by inter-channel scattering. }

A sharper potential near the edge  and narrower (or fully suppressed) compressible stripes in G/SiC would make  equilibration\cite{EdgeEquilibration, Kovrizhin, PartialEquilibration09, EdgeEquilibration14}  of edge channels
faster than in GraFETs with similar parameters. At the same time, cooling of edge state electrons by phonon emission\cite{SSFalko} would be slower in G/SiC, with hot electrons spreading to longer distances along the edge.

{\it Acknowledegements} We acknowledge useful discussions with K. von Klitzing, S.~Rozhko, A.~Tzalenchuk, J.T.~Janssen, R.~Nicholas, F.~Essler, J.~Wallbank, M.~Ben Shalom and constructive remarks of the Referees.  
This work was funded by Innovate UK grant 65431-468182 and the European Graphene Flagship project.

\end{document}